\documentclass[twocolumn,showpacs,preprintnumbers,amsmath,amssymb]{revtex4}

\usepackage{graphicx}
\usepackage{dcolumn}
\usepackage{amsfonts,amsmath,amssymb,bm}
\begin{document}

\title{Exchange parameters from approximate self-interaction correction scheme}

\author{A. Akande and S. Sanvito \thanks{sanvitos@tcd.ie}}
\affiliation{School of Physics and CRANN, Trinity College, Dublin 2, Ireland}

\date{\today}

\begin{abstract}
The approximate atomic self-interaction corrections (ASIC) method to density functional theory  is put to the 
test by calculating the exchange interaction for a number of prototypical materials, critical to local exchange and 
correlation functionals. ASIC total energy calculations are mapped onto an Heisenberg pair-wise interaction and 
the exchange constants $J$ are compared to those obtained with other methods. 
In general the ASIC scheme drastically improves the bandstructure, which for almost
all the cases investigated resemble closely available photo-emission data. In contrast the results for the exchange 
parameters are less satisfactory. Although ASIC performs reasonably well for systems where the magnetism originates
from half-filled bands, it suffers from similar problems than those of LDA for other situations. In particular the 
exchange constants are still overestimated. This reflects a subtle interplay between exchange and 
correlation energy, not captured by the ASIC.
\end{abstract}

\pacs{}
\keywords{}

\maketitle

\section{Introduction}

Theoretical studies based on density functional theory (DFT) \cite{akindft1,akindft2} have given remarkable insights 
into the electronic and magnetic properties of both molecules and solids \cite{akin1}. In particular, a 
number of these studies attempt to quantitatively describe the magnetic interaction in a broad range of systems including 
transition metals \cite{turek}, hypothetical atomic chains \cite{akin5,akin6}, ionic solids \cite{akin17,akin18,akin20}, 
transition metal oxides \cite{akin10,akin11} and transition metal polynuclear complexes \cite{akin2,akin13,akin14}. 
DFT uses an effective single-particle picture where spin symmetry is generally broken. For this reason exchange 
parameters $J$ are conventionally extracted by using a mapping procedure, where total energy calculations are 
fitted to a classical Heisenberg Hamiltonian \cite{turek,akin15}. This is then used for evaluating the Curie or N\'eel 
temperatures, the magnetic susceptibility and for interpreting neutron diffraction experiments \cite{akin15a}. 

Notably, the accuracy and reliability of the numerical values of the $J$'s depend on the functional used for the 
exchange and correlation (XC) energy, being this the only approximated part of the DFT total energy \cite{akin16}. 
Calculations based on well-known local functionals, namely the local density approximation (LDA) and the generalised 
gradient approximation (GGA), are successful with itinerant magnetism in transition metals \cite{turek}, but largely
over-estimates the Heisenberg exchange parameters in many other situations \cite{akin17, akin18,akin20,akin11,akin14}.
Additional corrections based on the kinetic energy density (metaGGA) \cite{metaGGA} marginally improves the agreement with
experiments \cite{akin20}, although an extensive investigation over several solid state systems has not been carried
out so far. 

These failures are usually connected to the local character of the LDA, which is only weakly modified by constructing
XC potentials including the gradient, or higher derivative of the charge density. A direct consequence is that the charge 
density is artificially delocalized in space, leading to an erroneous alignment of the magnetic bands. 
These are also artificially broadened. 
A typical example is that of NiO, which LDA predicts as Mott-Hubbard instead of charge-transfer insulator. 
Clearly a qualitative failure in describing the ground state results in an erroneous prediction of the exchange parameters. 

One of the reasons behind the inability of LDA and GGA of describing localized charge densities is attributed to
the presence of the self-interaction error (SIE) \cite{akin21}. This originates from the spurious Coulomb interaction of an electron with itself,
which is inherent to local functionals. Hartree-Fock (HF) methods, in the unrestricted or spin polarised form, are SIE free
and produce systematic improved $J$ parameters. However, these methods lack of correlation and usually overcorrect.
A typical example is the antiferromagnetic insulator KNiF$_3$ for which HF predicts a nearest neighbour $J$ of 
around 2~meV \cite{akin17,
akin3,akin4,akin8,akin9} against an experimental value of 8.6~meV \cite{akin28}. Direct SIE subtraction, conventionally
called self-interaction corrected (SIC) LDA, also improves the results and seems to be less affected by overcorrection
\cite{akin5,temm}. Similarly hybrid-functionals, which mix portions of HF exchange with the local density
approximation of DFT, perform better than local functionals and in several situations return values for $J$ in close
agreement with experiments \cite{akin17,akin18}. 

It is important to note that both methods based non-local exchange or SIC, are computationally demanding and thus their 
application to the solid state remains rather limited. It is then crucial to develop practical computational schemes able to provide
a good estimate of the exchange parameters for those systems critical to LDA, which at the same time are not
numerically intensive. Based on the idea that most of the SIE originates from highly localized states, with a charge
distribution resembling those of free atoms, Vogel et al. \cite{akin20e} proposed a simple SIC scheme where 
the corrections are approximated by a simple on-site term. This method was then generalized to fractional occupation by
Filippetti and Spaldin \cite{akin20f} and then implemented in a localized atomic orbital code for large scaling by
Pemmaraju et al. \cite{akin22}. Despite its simplicity the method has been successfully applied to a number of interesting 
physical systems including , transition metal monoxides \cite{akin20f,alessio2},
silver halides \cite{voeg3}, noble metal oxides \cite{alessio3}, 
ferroelectric materials \cite{akin20f,alessio4,alessio5}, high-k materials \cite{alessio6}, diluted magnetic 
semiconductors \cite{FSS1,FSS2} and also to quantum transport \cite{cormac1,cormac2}. 

The method is strictly speaking not variational, in the sense that a functional generating the ASIC potential 
via variational principle is not available. However, since typically the LDA energy is a good approximation of the exact DFT 
energy, although the LDA potential is rather different from the exact KS potential, a ``practical'' definition of total 
energy can be provided. In this work we evaluate the ability of this approximated energy in describing exchange
parameters for a variety of magnetic systems. 


\section{The atomic SIC method}

The seminal work of Perdew and Zunger \cite{akin21} pioneered the modern theory of SIC. The main idea is that 
of subtracting directly the spurious SI for each Kohn-Sham (KS) orbital $\psi_n$. The SIC-LDA \cite{note} XC
energy thus writes
\begin{equation}\label{p1}
E_\mathrm{xc}^\mathrm{SIC}[\rho^{\uparrow},\rho^{\downarrow}]=
E_\mathrm{xc}^\mathrm{LDA}[\rho^{\uparrow},\rho^{\downarrow}]-\sum_{n\sigma}^\mathrm{occupied}\delta_n^\mathrm{SIC} ,
\end{equation}
where $E_\mathrm{xc}^\mathrm{LDA}[\rho^{\uparrow},\rho^{\downarrow}]$ is the LDA-XC energy and
$\delta_n^\mathrm{SIC}$ is the sum of the self-Hartree and self-XC energy associated to the charge 
density $\rho_n^{\sigma}=|\psi_n^\sigma|^2$ of the fully occupied KS orbital $\psi_n^{\sigma}$ 
\begin{equation}\label{p2}
\delta_n^\mathrm{SIC}=U[\rho_n^{\sigma}] + E_\mathrm{xc}^\mathrm{LDA}[\rho_n^{\sigma},0]\:.
\end{equation}
Here $U$ is the Hartree energy and $\sigma$ is the spin index.

The search for the energy minimum is not trivial, since $E_\mathrm{xc}^\mathrm{SIC}$ 
is not invariant under unitary rotations of the occupied KS orbitals. As a consequence the KS
method becomes either non-orthogonal or size-inconsistent. These problems however can be avoided \cite{lin1,lin2,lin3} 
by introducing a second set of orbitals $\phi_n^\sigma$ related to the canonical KS orbitals by a unitary transformation ${\cal M}$
\begin{equation}\label{unit}
\psi_n^\sigma=\sum_m{\cal M}_{nm}^\sigma\phi_m^\sigma\:.
\end{equation}

The functional can then be minimized by varying both the orbitals $\psi$ and the unitary transformation ${\cal M}$,
leading to a system of equations
\begin{equation}
H_n^\sigma\psi_n^\sigma=(H_0^\sigma+\Delta v^\mathrm{SIC}_n)\psi_n^\sigma=
\epsilon_{n}^{\sigma,\mathrm{SIC}}\psi_n^\sigma\;,
\label{KSSIC-pro}
\end{equation}
\begin{equation}
\psi^\sigma_n=\sum_m{\cal M}_{nm}\phi_m^\sigma\;,
\label{SIC-Rot}
\end{equation}
\begin{equation}
\Delta v^\mathrm{SIC}_n=\sum_m{\cal M}_{nm}v^\mathrm{SIC}_m\frac{\phi_m^\sigma}{\psi_n^\sigma}=
\sum_m v^\mathrm{SIC}_m \hat{P}_m^\phi\;,
\label{DSIC}
\end{equation}
where $H_0^\sigma$ is the LDA Hamiltonian, 
$\hat{P}_m^\phi\psi_n^\sigma({\bf r})=\phi_m^\sigma({\bf r})\langle\phi_m^\sigma|\psi_n^\sigma\rangle$ and
$v^\mathrm{SIC}_n=u([\rho_n]; {\bf r})+v_\mathrm{xc}^{\sigma,\mathrm{LDA}}([\rho_n^\uparrow,0]; {\bf r})$,
with $u$ and $v_\mathrm{xc}^{\sigma,\mathrm{LDA}}$ the Hatree and LDA-XC potential respectively.

In equation (\ref{KSSIC-pro}) we have used the
fact that at the energy minimum the matrix of SIC KS-eigenvalues $\epsilon_{nm}^{\sigma,\mathrm{SIC}}$ is
diagonalized by the KS orbitals $\psi_n$. Importantly such minimization scheme can be readily applied to
extended systems, without loosing the Bloch representation of the KS orbitals \cite{akin20c,temm1}. 

The ASIC method consists in taking two drastic approximations in equation (\ref{KSSIC-pro}). First
we assume that the orbitals $\phi_m$, that minimize the SIC functional are atomic-like orbitals 
$\Phi_m^\sigma$ (ASIC orbitals) thus
\begin{equation}
\sum_m v^\mathrm{SIC}_m({\bf r}) \hat{P}_m^\phi\:\rightarrow\:
\alpha\:\sum_m \tilde{v}^{\sigma\mathrm{SIC}}_m({\bf r}) \hat{P}_m^\Phi
\;,
\label{PSIC-appr}
\end{equation}
where $\tilde{v}^{\sigma\mathrm{SIC}}_m({\bf r})$ and $\hat{P}_m^\Phi$ are the SIC potential and the projector
associated to the atomic orbital $\Phi_m^\sigma$. Secondly we replace the non-local projector $\hat{P}_m^\Phi$ 
with its expectation value in such a way that the final ASIC potential reads
\begin{equation}
v_\mathrm{ASIC}^\sigma({\bf r})=\alpha\:\sum_m \tilde{v}^{\sigma\mathrm{SIC}}_m({\bf r}) p_m^\sigma\;,
\label{PSIC-appr-final}
\end{equation}
where $p_m^\sigma$ is the orbital occupation (essentially the spin-resolved M\"ulliken orbital 
population) of $\Phi_m$.

Note that in the final expression for the potential a factor $\alpha$ appears. This is an empirical scaling term
that accounts for the fact that the ASIC orbital $\Phi$ in general do not coincide with those that minimize the SIC functional
(\ref{p1}). By construction $\alpha=1$ in the single particle limit, while it vanishes for the homogeneous electron gas. 
Although in general $0<\alpha<1$, extensive testing \cite{akin22} demonstrates that a value around 1 describes well
ionic solids and molecules, while a value around 1/2 is enough for mid- to wide-gap insulators. In the following
we will label with ASIC$_{1/2}$ and ASIC$_{1}$ calculations obtained respectively
with $\alpha=1/2$ and $\alpha=1$.

Finally we make a few comments over the total energy. As pointed out in the introduction the present theory is not
variational since the KS potential cannot be related to a functional by a variational principle. However, since
typical LDA energies are more accurate than their corresponding KS potentials, we use the expression of
equation (\ref{p1}) as suitable energy. In this case the orbital densities entering the SIC are those given
by the ASIC orbital $\Phi$. Moreover, in presenting the data, we will distinguish results obtained by using 
the SIC energy (\ref{p1}) from those obtained simply from the LDA functional evaluated at the ASIC density,
i.e. without including the $\delta_n$ corrections (\ref{p2}).


\section{Results}

All our results have been obtained with an implementation of the ASIC method \cite{akin22} based on
the DFT code {\it Siesta} \cite{akin23}. {\it Siesta} is an advanced DFT code using pseudopotentials and
an efficient numerical atomic orbital basis set. In order to compare the exchange parameters obtained with different
XC functionals we consider the LDA parameterization of Ceperly and Alder \cite{akin24}, the GGA functional obtained by 
combining Becke exchange \cite{akin25} with Lee-Yang-Parr correlation \cite{akin25B} (BLYP), the nonempirical Purdew, 
Burke and Ernzerhof (PBE) GGA \cite{akin26}, and the ASIC scheme as implemented in reference \cite{akin22}.

Calculations are performed for different systems critical to LDA and GGA, ranging from molecules to extended
solids. These include hypothetical H-He atomic chains, the ionic solid KNiF$_3$ and the transition metal monoxides 
MnO and NiO. DFT total energy calculations are mapped onto an effective pairwise Heisenberg Hamiltonian
\begin{equation}
H_\mathrm{H}=-\sum_{\langle nm\rangle}J_{nm}\vec{S}_n\cdot\vec{S}_m\:,
\end{equation}
where the sums runs over all the possible pairs of spins. In doing this we wish to stress that the mapping is a 
convenient way of comparing total energies of different magnetic configurations calculated with different
functionals. In this spirit the controversy around using the spin-projected (Heisenberg mapping) or the 
non-projected scheme is immaterial \cite{akin5,pole1,pole2}.

\subsection{H-He chain}
\begin{figure}[htb]
\begin{center}
\includegraphics[width=0.35\textwidth]{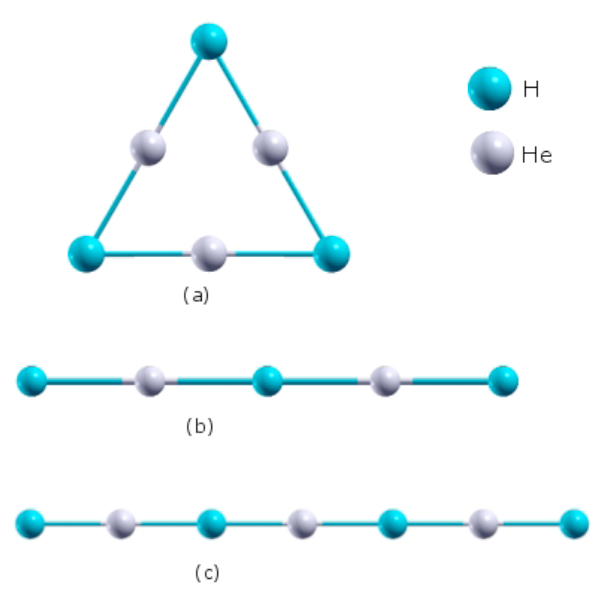}
\end{center}
\caption[] {(Color on-line) H-He-H chains at an inter-atomic distance of 1.625\AA.} \label{chain}
\end{figure}

As an example of molecular systems, we consider H-He monoatomic chains at a 
inter-atomic separation of 1.625~\AA\ (see figure \ref{chain}). This is an important benchmark for DFT since 
the wave-function is expected to be rather localized and therefore to be badly described by local
XC functionals. In addition the system is simple enough to be accessible by accurate quantum chemistry 
calculations. 

As basis set we use two radial functions (double-$\zeta$) for the $s$ and $p$ angular momenta of both H and He, while
the density of the real-space grid converges the self-consistent calculation at 300 Ry.
Here we consider all possible Heisenberg parameters. Thus the triangular molecule (Fig.\ref{chain}a) has only one 
nearest neighbour parameter $J_{12}^a$, the 5-atom chain (Fig.\ref{chain}b) has both first $J_{12}^b$ and second 
neighbour $J_{13}^b$ parameters, and the 7-atom chain (Fig.\ref{chain}c) has three parameters describing
respectively the nearest neighbour interaction with peripheral atoms $J_{12}^c$, the nearest neighbour interaction 
between the two middle atoms $J_{23}^c$ and the second neighbour interaction $J_{13}^c$.

Following reference \cite{akin5}, accurate calculations based on second-order perturbation theory (CASPT2) \cite{akin2} are 
used as comparison. The quality of each particular functionals is measured as the relative mean deviation 
of the nearest neighbour exchange parameters only ($J_{12}^a$, $J_{12}^b$, $J_{12}^c$, $J_{23}^a$), since those are 
the largest ones
\begin{equation}
\delta={1 \over 4} \sum_i^4 \frac{|J_i - J^\mathrm{CASPT2}_i|}{|J^\mathrm{CASPT2}_i|}\:.
\end{equation}

Our calculated $J$ values and their relative $\delta$ are presented in table \ref{table1}, where we also include
results for a fully self-consistent SIC calculation over the B3LYP functional (SIC-B3LYP) \cite{akin5}.
\begin{table}
\begin{center}
\begin{tabular}{lccccccc}
\hline\hline
Method & $J_{12}^a$ & $J_{12}^b$ & $J_{13}^b$ & $J_{12}^c$ & $J_{23}^c$ & $J_{13}^c$ & $\delta$ (\%)\\
\hline\hline
 CASPT2 &-24 & -74 & -0.7 & -74 & -79 & -0.7  & 0\\
\hline
 SIC-B3LYP & -31 &-83 & -0.2 & -83 & -88 & -0.3  & 16 \\
\hline
 LDA &  -68 &-232 & -6 & -234 & -260 & -6 & 210 \\
\hline
 PBE &  -60 & -190 & -1.8 & -190 & -194 & -1.6 & 152 \\  
\hline
 BLYP & -62 &-186 & -2 &  -186 & -200 & -1  & 147 \\
\hline
 ASIC$_{1}$ & -36 &-112 & -1  & -110 & -122 & -0.6 & 51  \\
\hline
 ASIC$_{1/2}$ & -44 & -152 & -1 & -152 & -168 & -1.4 & 101  \\
\hline
 ASIC$_{1}^*$ & -40 &-128 & -0.6  & -128 & -142 & -1.0 & 73  \\
\hline
 ASIC$_{1/2}^*$ & -50 & -170 & -1.4 & -170 & -190 & -1.8 & 127  \\
\hline\hline
\end{tabular}
\end{center}
\caption{Calculated $J$ values (in meV) for the three different H--He chains shown in Fig.\ref{chain}. The
CASPT2 values are from reference \cite{akin2}, while the SIC-B3LYP are from reference \cite{akin5}. The last
two rows correspond to $J$ values obtained from the LDA energy calculated at the ASIC density.}
\label{table1}
\end{table}
It comes without big surprise that the LDA systematically overestimates all the exchange parameters with errors up 
to a factor 6 for the smaller $J$ ($J_{13}^b$ and $J_{13}^c$) and an average error $\delta$ for the largest $J$ of about 200\%.
Standard GGA corrections considerably improve the description although the $J$'s are still systematically
larger than those obtained with CASPT2. Note that the results seem rather independent of the particular GGA
parameterization, with PBE and BLYB producing similar exchange constants. This is in good agreement with
previous calculations \cite{akin5}.

SIC in general dramatically improves the LDA and GGA description and our results for ASIC$_{1}$ are 
reasonably close to those obtained with the full self-consistent procedure (SIC-B3LYP). This is an interesting
result, considering that our ASIC starts from a local exchange functional, while B3LYP already contains
non-local contributions. We also evaluate the $J$ parameters by using the LDA energy evaluated at the ASIC density 
(last two rows in table \ref{table1}). In general this procedure gives $J$'s larger than those obtained by using the 
energy of equation (\ref{p1}), meaning that the $\delta_n^\mathrm{SIC}$ contributions reduce the $J$ values. 

It is then clear that the ASIC scheme systematically improves the $J$ values as compared to local functionals. 
The agreement however is not as good as the one obtained by using a fully self-consistent SIC scheme, meaning that
for this molecular system the ASIC orbitals are probably still not localized enough. This can alter the actual contribution of 
$\delta_n^\mathrm{SIC}$ to the total energy and therefore the exchange parameters. 

\subsection{Ionic antiferromagnets: KNiF$_3$}

\begin{figure}[htb]
\begin{center}
\includegraphics[width=0.25\textwidth,angle=0.0]{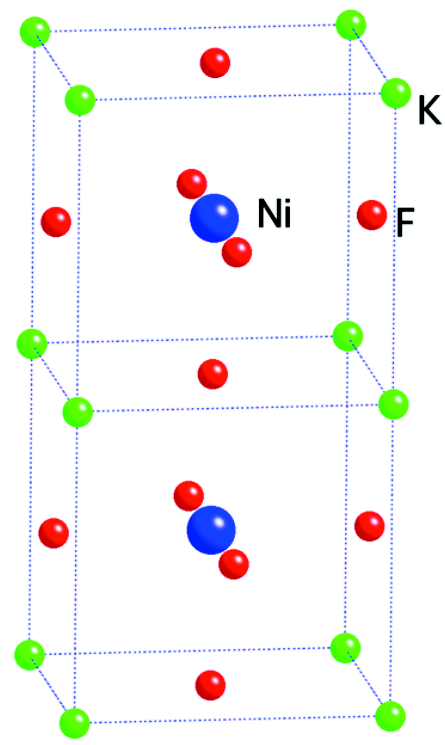}
\end{center}
\caption[]{(Color on-line) Cubic perovskite structure of KNiF$_3$. Color code: blue=Ni, red=F, Green=K.} \label{cube}
\end{figure}

Motivated by the substantial improvement of ASIC over LDA, we then investigate 
its performances for real solid-state systems, starting from KNiF$_3$. 
This is a prototypical Heisenberg antiferromagnet with strong ionic character, a material for which our
ASIC approximation is expected to work rather well \cite{akin22}. It is also a well studied material, both
experimentally \cite{akin28,akin29} and theoretically \cite{akin4,akin8,akin17,akin20}, allowing us extensive
comparisons. The KNiF$_3$ has cubic perovskite-like structure with the nickel atoms at the edges of the cube, 
flourine atoms at the sides and potassium atoms at the center (see Fig.\ref{cube}). At low temperature, KNiF$_3$ 
is a type II antiferromagnetic insulator consisting of ferromagnetic (111) Ni planes aligned antiparallel to each other.
For our calculations we use a double-$\zeta$ polarized basis for the $s$ and $p$ orbitals of K, Ni and F,
a double-$\zeta$ for the 3$d$ of K and Ni, and a single-$\zeta$ for the 3$d$ of F. Finally, we use 5$\times$5$\times$5 
$k$-points in the full Brillouin zone and the real-space mesh cutoff is 550 Ry. Note that the configuration
used to generate the pseudopotential is that of Ni$^{2+}$, 4$s^1$3$d^7$.

\begin{figure}[htb]
\begin{center}
\includegraphics[width=0.45\textwidth,angle=0.0]{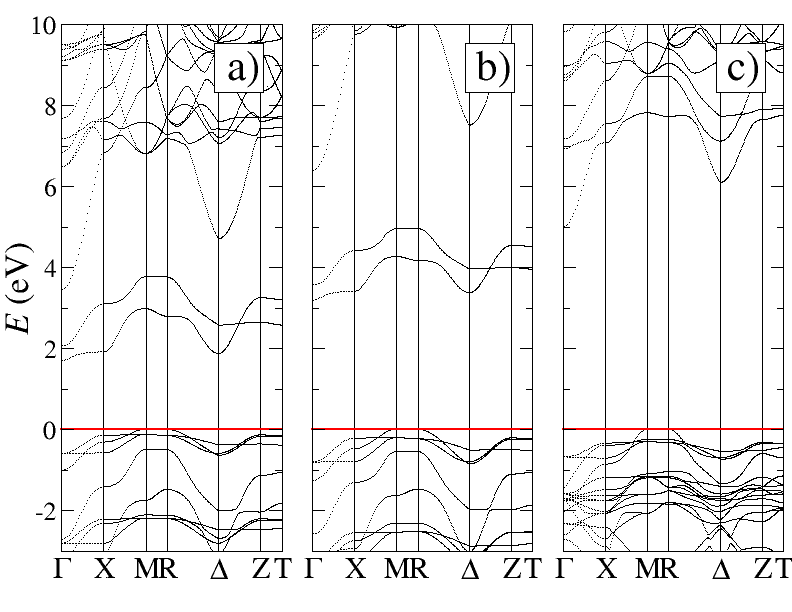}
\end{center}
\caption[] {Band structure for type II antiferromagnetic KNiF$_3$ obtained with a) LDA, b) ASIC$_1$ and c) LDA+$U$ 
($U$=8~eV and $J$=1~eV). The valence band top is aligned at $E$=$E_\mathrm{F}$=0~eV (horizontal line).} \label{knifband}
\end{figure}

\begin{figure}[htb]
\begin{center}
\includegraphics[width=0.4\textwidth,angle=0.0]{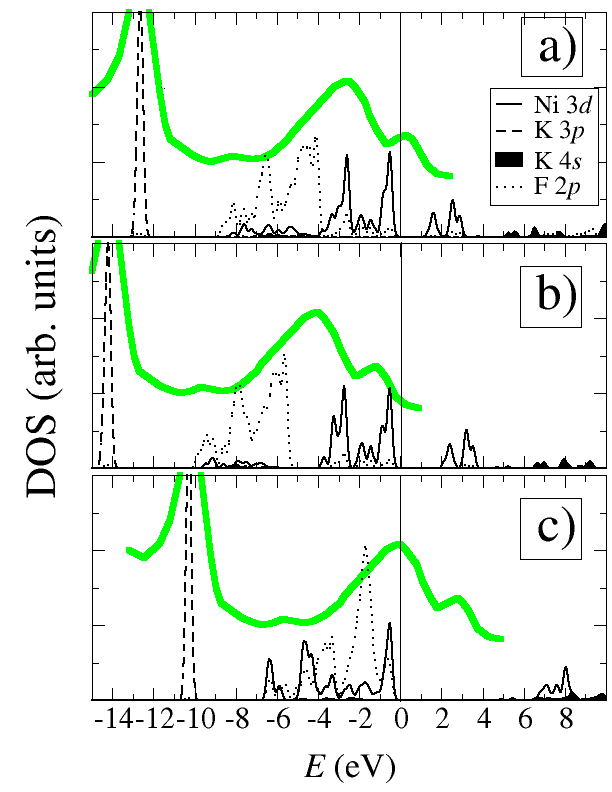}
\end{center}
\caption[] {(Color on-line) DOS for type II antiferromagnetic KNiF$_3$ obtained with a) LDA, b) ASIC$_1$ and c) LDA+$U$ 
($U$=8~eV and $J$=1~eV). The valence band top is aligned at $E$=0~eV (vertical line). The experimental UPS spectrum
from reference \cite{onuki} is also presented (thick green line). The relative binding energy is shifted in order to match the
K 3$p$ peak.} \label{knifdos}
\end{figure}

We first consider the band-structure as obtained with LDA and ASIC. For comparison we also include results
obtained with LDA+$U$ \cite{akin29a,akin29b} as implemented in {\it Siesta} \cite{gosia}. In this case we correct
only the Ni $d$ shell and we fix the Hubbard-$U$ and Hund's exchange-$J$ parameters by fitting the 
experimental lattice constant ($a_0=4.014$~\AA). 
The calculated values are $U$=8~eV and $J$=1~eV. The bands obtained with
the three methods and the corresponding orbital projected density of states (DOS) are presented in figures
\ref{knifband} and \ref{knifdos} respectively.

All the three functionals describe KNiF$_3$ as an insulator with bandgaps respectively of 1.68~eV (LDA), 3.19~eV (ASIC$_1$),
and 5.0~eV (LDA+$U$). An experimental value for the gap is not available and therefore a comparison cannot be made.
In the case of LDA and ASIC the gap is formed between Ni states, with conductance band bottom well described by
$e_g$ orbitals. These are progressively moved upwards in energy by the SIC, but still occupy the gap. Such feature is
modified by LDA+$U$ which pushes the unoccupied $e_g$ states above the conductance band minimum, which is now
dominated by K 4$s$ orbitals. 

In more detail the valence band is characterized by a low-lying K 3$p$ band and by a mixed Ni-3$d$/F 2$p$. While the
K 3$p$ band is extremely localized and does not present substantial additional orbital components the amount of mixing 
and the broadening of the Ni-3$d$/F 2$p$ varies with the functionals used. In particular both LDA and ASIC predict that
the Ni 3$d$ component occupies the high energy part of the band, while the F 2$p$ the lower. For both the total bandwidth
is rather similar and it is about 9-10~eV. In contrast LDA+$U$ offers a picture where the Ni-F hybridization spread across 
the whole bandwidth, which is now reduced to less than 7~eV.

Experimentally, ultraviolet photoemission spectroscopy (UPS) study of the whole KMF$_3$ (M: Mn, Fe, Co, Ni, Cu, Zn) 
series \cite{onuki} gives 
us a spectrum dominated by two main peaks: a low K 3$p$ peak and broad band mainly attributed to F 2$p$. These two
spectroscopical features are separated by a binding energy of about 10~eV. In addition the 10~eV wide F 2$p$ band has
some fine structure related to various Ni $d$ multiplets. An analysis based on averaging the multiplet structure \cite{onuki} 
locates the occupied Ni $d$ states at a bounding energy about 3~eV smaller than that of the F 2$p$ band. In figure 
\ref{knifdos} we superimpose the experimental UPS spectrum to our calculated DOS, with the convention of aligning in
each case the sharp K 3$p$ peak.

It is then clear that ASIC provides in general a better agreement with the UPS data. In particular both the Ni-3$d$/F 2$p$
bandwidth and the position of the Fermi energy ($E_\mathrm{F}$) with respect to the K 3$p$ peak are correctly predicted. 
This is an improvement over LDA, which describes well the Ni-3$d$/F 2$p$ band, but positions the K 3$p$ states too close 
to $E_\mathrm{F}$. For this reason, when we align the UPS spectrum at the K 3$p$ position, this extends over $E_\mathrm{F}$.
Finally in the case of LDA+$U$, there is a substantial misalignment between the UPS data and our DOS. LDA+$U$ in fact 
erroneously pushes part of the Ni $d$ mainfold below the F 2$p$ DOS, which now forms a rather narrow band. 

We now turn our attention to total energy related quantities. In table \ref{table2} we present the theoretical equilibrium
lattice constant $a_0$ and the Heisenberg exchange parameter $J$ for all the functionals used. Experimentally we have 
$J$=8.2$\:\pm$~0.6~meV \cite{akin28}. The values of $a_0$ and $J$ are calculated for the type II
antiferromagnetic ground state, by constructing a supercell along the (111) direction. Importantly values of $J$ 
obtained by considering a supercell along the (100) direction, i.e. by imposing antiferromagnetic alignment between ferromagnetic 
(100) planes (type I antiferromagnet), yield essentially the same result, confirming the fact that the interaction is effectively only extending to
nearest neighbors. Furthermore we report results obtained both at the theoretical equilibrium lattice constant ($J_\mathrm{th}$)
and at the experimental one ($J_\mathrm{ex}$).
\begin{table}
\begin{center}
\begin{tabular}{llllll}
\hline\hline
 Method & $a_0$ & $J_\mathrm{th}$ & $P_d^\mathrm{th}$ & $J_\mathrm{ex}$ & $P_d^\mathrm{ex}$ \\
\hline 
 LDA & 3.951 & 46.12 (53.1) & 1.829 & 40.4 & 1.834 \\
\hline
 PBE & 4.052 & 33.98 (37.0) & 1.813 & 36.48 & 1.808  \\
\hline
 BLYP & 4.091 & 31.10 (37.6) & 1.821 & 36.72 & 1.812 \\
\hline
 ASIC$_{1/2}$ & 3.960 & 40.83 & 1.876 & 36.14  & 1.878 \\ 
\hline
 ASIC$_1$  & 3.949 & 36.22 & 1.907 & 30.45  &  1.914 \\
\hline
 ASIC$_{1/2}^*$ & 3.969 & 43.44 & 1.876 & 38.57  & 1.878 \\ 
\hline
 ASIC$_1^*$  & 3.949 & 39.80 & 1.907 & 33.56  &  1.914 \\
\hline
 LDA+U  & 4.007 & 12.55 & & 10.47 & 1.940     \\
\hline\hline
\end{tabular}
\end{center}
\caption{Calculated $J$ parameters (in meV) and the M\"ulliken magnetic moment for Ni 3$d$ ($P_d$) in KNiF$_3$. 
The experimental values for $J$ and $a_0$ are 8.2$\:\pm$0.6~meV and 4.014\AA\ respectively while the values 
in brackets are those from reference \cite{akin20}. In the table we report values evaluated at the 
theoretical ($J_\mathrm{th}$ and $P_d^\mathrm{th}$) and experimental ($J_\mathrm{ex}$ and $P_d^\mathrm{ex}$) 
lattice constant. ASIC$_{1/2}^*$ and ASIC$_{1}^*$ are obtained from the LDA energies evaluated at the ASIC density.}\label{table2}
\end{table}

Also in this case local XC functionals largely overestimate $J$, with errors for $J_\mathrm{ex}$ going from a factor 8 
(LDA) to a factor 4.5 (GGA-type). ASIC improves these values, although only marginally, and our best agreement is 
found for ASIC$_1$, while ASIC$_{1/2}$ is substantially identical to GGA. Interestingly the ASIC$_1$ performance is 
rather similar, if not better, to that of meta-GGA functionals \cite{akin20}. The situation is however worsened when we 
consider $J$ parameters obtained at the theoretical lattice constant. The ASIC-calculated $a_0$ are essentially 
identical to those from LDA and about 2\% shorter than those from GGA. Since $J$ depends rather severely on
the lattice parameter we find that at the theoretical lattice constant GGA-functionals perform actually better than
our ASIC. Finally, also in this case the $J$'s obtained by simply using the LDA energies are larger than those
calculated by including the SIC corrections (see equation \ref{p1}).

In general the improvement of the $J$ parameter is correlated to an higher degree of electron localization, 
in particular of the Ni $d$ shell. In table \ref{table2} the magnetic moment of the Ni $d$ shell $P_d$, obtained from the 
M\"ulliken population, is reported. This increases systematically when going from LDA to GGA to ASIC approaching the
atomic value expected from Ni$^{2+}$. 

Our best result is obtained with LDA+$U$, which returns an exchange of 10.47~meV for the same $U$ and $J$ 
that fit the experimental lattice constant. This is somehow superior performance of LDA+$U$ with respect to ASIC 
should not be surprising and it is partially related to an increased localization. 
The Ni ions $d$ shell in octahedral coordination splits into $t_{2g}$ and $e_g$ states,
which further split according to Hund's rule. The $t_{2g}$ states are all filled, while for the $e_g$ only 
the majority are. By looking at the LDA DOS one can recognize the occupied $t_{2g}^\uparrow$ orbitals 
(we indicate majority and minority spins respectively with $\uparrow$ and $\downarrow$) at -3~eV, 
the $e_g^\uparrow$ at -2~eV and the $t_{2g}^\downarrow$ at about 0~eV, while the empty $e_g^\downarrow$ are
at between 1 and 3~eV above the valence band maximum. 

The local Hund's split can be estimated from the $e_g^\uparrow$-$e_g^\downarrow$ separation. The ASIC 
scheme corrects only occupied states \cite{note2}, and therefore it enhances the local exchange by only a 
downshift of the valence band. From the DOS of figure \ref{knifdos} it is clear that this is only a small contribution. 
In contrast the LDA+$U$ scheme also corrects empty states, effectively pushing upwards in energy the 
$e_g^\downarrow$ band. The net result is that of a much higher degree of localization of the $d$ shell with a 
consequent reduction of the Ni-Ni exchange. This is similar to the situation described by the Hartree-Fock method, 
which however returns exchange parameters considerably smaller than the experimental value \cite{akin3,akin4,akin8,akin9}. 
Interestingly hybrid functionals \cite{akin17} have the right mixture of non-local exchange and electron correlation
and produce $J$'s in close agreement with the experiments. 

We further investigate the magnetic interaction by evaluating $J$ as a function of the lattice constant. Experimentally
this can be achieved by replacing K with Rb and Tl, and indeed de Jongh and Block \cite{JoBlo} early suggested a
$d^{-\alpha}$ power law with $\alpha=12\pm2$. Our calculated $J$ as a function of the
lattice constant $d$ for LDA, GGA, ASIC$_1$ and LDA+$U$ ($U$=8~eV and $J$=1~eV) are presented in figure \ref{Fig5.png}.
\begin{figure}[htb]
\begin{center}
\includegraphics[width=0.4\textwidth,angle=0.0]{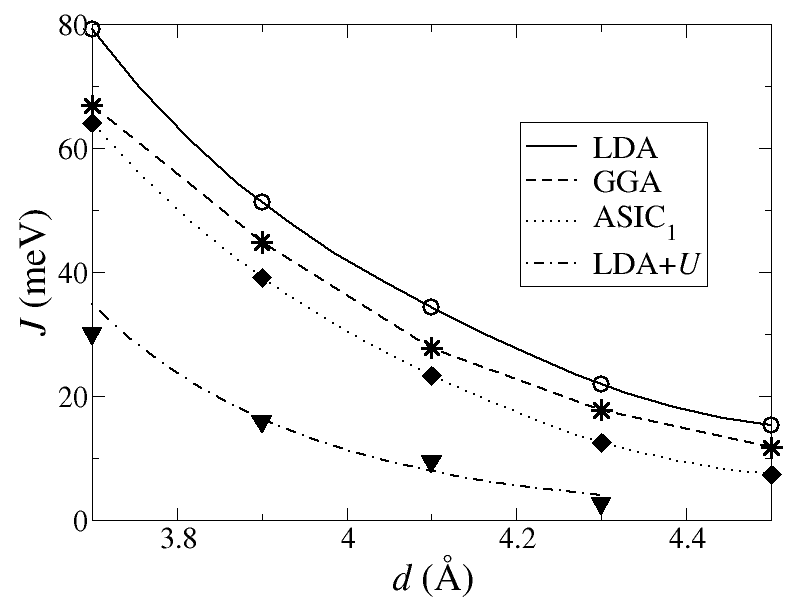}
\end{center}
\caption[] {$J$ as a function of the lattice constant for LDA, GGA, ASIC$_1$ and LDA+$U$ ($U$=8~eV and $J$=1~eV).
The symbols are our calculate value while the solid lines represent the best power-law fit.} \label{Fig5}
\end{figure}
For all the four functionals investigated $J$ varies as a power law, although the calculated exponents are rather different:
8.6 for LDA, 9.1 for GGA, 11.3 for ASIC$_1$ and 14.4 for LDA+$U$. This further confirms the strong underestimation of
the exchange constants from local functionals. Clearly the relative difference between the $J$ obtained with different 
functionals becomes less pronounced for small $d$, where the hybridization increases and local functionals perform better. 
Note that only ASIC$_1$ is compatible with the experimental exponent of $12\pm2$, being the one evaluated from LDA+$U$
too large. Importantly we do not expect to extrapolate the LDA+$U$ value at any distance, since the screening 
of the parameters $U$ and $J$ changes with the lattice constant. 

In conclusion for the critical case of KNiF$_3$ the ASIC method appears to improve the LDA results. This is 
essentially due to the better degree of localization achieved by the ASIC as compared with standard local
functionals. However, while the improvement over the bandstructure is substantial, it is only marginal for
energy-related quantities. The main contribution to the total energy in the ASIC scheme comes from the LDA
functional, which is now evaluated at the ASIC density. This is not sufficient for improving the exchange parameters,
which in contrast need at least a portion of non-local exchange.

\subsection{Transition metal monoxides}

Another important test for the ASIC method is that of transition metal monoxides. These have been extensively studied both 
experimentally and theoretically and they are the prototypical materials for which the LDA appears completely inadequate. In this work 
we consider MnO and NiO, which have respectively half-filled and partially-filled 3$d$ shells. They both crystallize in the rock-salt structure 
and in the ground state they are both type-II antiferromagnetic insulators. 
The N\'eel's temperatures are 116~K and 525~K respectively for MnO and NiO. In all our calculations we consider 
double-$\zeta$ polarised basis for the $s$ and $p$ shell of Ni, Mn and O, double-$\zeta$ for the Ni and Mn 3$d$ orbitals, and single-$\zeta$ 
for the empty 3$d$ of O. We sample 6$\times$6$\times$6 $k$-points in the full Brillouin zone of both the cubic and rhombohedral 
cell describing respectively type I and type II antiferromagnetism. Finally the real-space mesh cutoff is 500~Ry.

\begin{figure}[htb]
\begin{center}
\includegraphics[width=0.48\textwidth,angle=0.0]{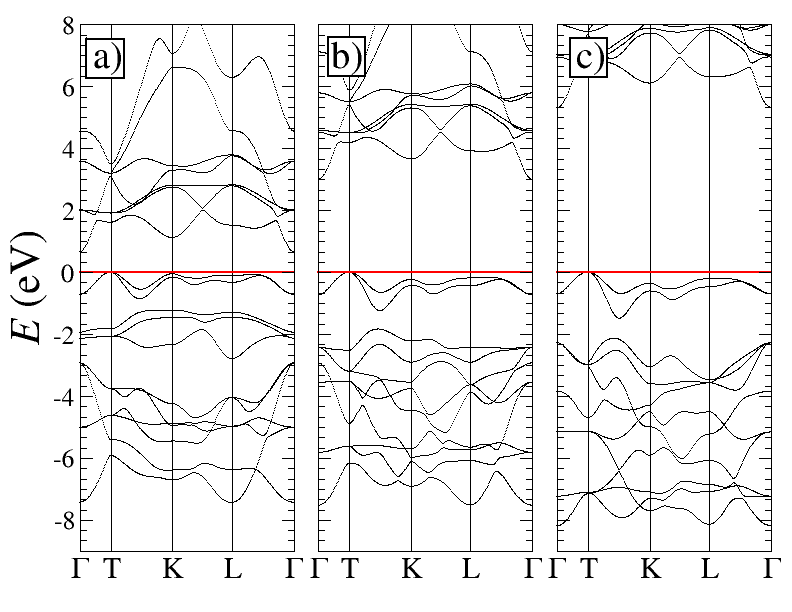}
\end{center}
\caption[] {Calculated band structure for the type II anti-ferromagnetic MnO obtained from a) LDA, 
b) ASIC$_{1/2}$ and c) ASIC$_1$. The valence band top is aligned at 0~eV (horizontal line).} \label{MnO}
\end{figure}

\begin{figure}[htb]
\begin{center}
\includegraphics[width=0.48\textwidth,angle=0.0]{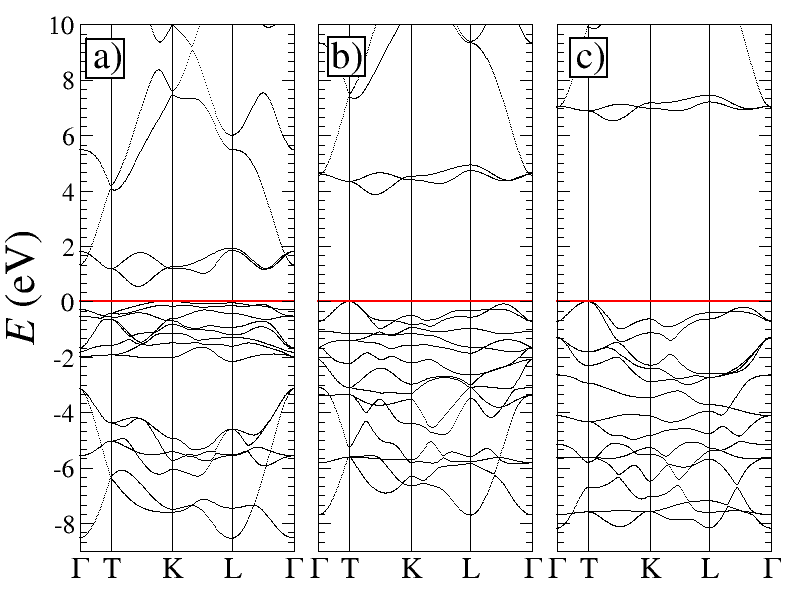}
\end{center}
\caption[] {Calculated band structure for the type II anti-ferromagnetic NiO obtained from a) LDA, 
b) ASIC$_{1/2}$ and c) ASIC$_1$. The valence band top is aligned at 0~eV (horizontal line).} \label{NiO}
\end{figure}

The calculated band structures obtained from LDA, ASIC$_{1/2}$ and ASIC$_{1}$ are shown in figures \ref{MnO} and \ref{NiO} for 
MnO and NiO respectively. 
These have been already discussed extensively in the context of the ASIC method \cite{akin20f,akin22} and here we report only the 
main features. For both the materials LDA already predicts an insulating behavior, although the calculated gaps are rather small and
the nature of the gaps is not what experimentally found. In both cases the valence band top has an almost pure $d$ component, which
suggests these materials to be small gap Mott-Hubbard insulators. The ASIC downshifts the occupied $d$ bands which now hybridize
with the O-$p$ manifold. The result is a systematic increase of the band-gap which is more pronounced as the parameter $\alpha$ goes
from $1/2$ to 1. Importantly, as noted already before \cite{akin22}, the experimental band-gap is obtained for $\alpha\sim1/2$.

We then moved to calculating the exchange parameters. In this case we extend the Heisenberg model to second nearest 
neighbors, by introducing the first ($J_1$) and second ($J_2$) neighbor exchange parameters. These are evaluated 
from total energy calculations for a ferromagnetic and both type II and type I antiferromagnetic alignments. Our calculated results, 
together with a few selected data available from the literature are presented in table \ref{table3}.
\begin{table}
\begin{ruledtabular}
\begin{tabular}{lccccccc}
Method&\multicolumn{3}{c}{MnO}& $\;\;\;$ &\multicolumn{3}{c}{NiO}\\
\cline{2-8}\\
& $J_1$   & $J_2$ & $P_d$ & & $J_1$ & $J_2$ & $P_d$ \\
\hline\hline
LDA &  1.0 & 2.5 & 4.49 (4.38) & & 13.0 & -94.4 & 1.41 (1.41) \\
\hline
PBE & 1.5 & 2.5 & 4.55 (4.57) & & 7.0 & -86.8 & 1.50 (1.59) \\
\hline
ASIC$_{1/2}$  & 1.15 & 2.44 & 4.72 (4.77) & & 6.5 & -67.3 & 1.72 (1.77) \\
\hline
ASIC$_1$  & 0.65 &  1.81 & 4.84 (4.86) & & 3.8 & -41.8 & 1.83 (1.84) \\
\hline
ASIC$_{1/2}^*$  & 1.27 & 2.65 & 4.72 (4.77) & & 7.1 & -74.6 & 1.72 (1.77) \\
\hline
ASIC$_1^*$  & 0.69 &  2.03 & 4.84 (4.86) & & 4.4 & -47.9 & 1.83 (1.84) \\
\hline
SE1$^a$ & 0.86 & 0.95 & & & & & \\
\hline
HF$^b$ & 0.22 & 0.36 & & & & & \\
\hline
B3LYP$^c$ & 0.81 & 1.71 & & & & & \\
\hline
PBE0$^b$ & 0.96 & 1.14 & & & & & \\
\hline
B3LYP$^d$& & & & & 2.4 &-26.7 & \\
\hline
HF$^d$ &  &  & & & 0.8 & -4.6 & \\
\hline
SIC-LDA$^e$ & & & & & 2.3 & -12 & \\
\hline
Expt.$^f$ & & & & & 1.4 & -19.8 \\
\hline
Expt.$^g$ & & & & & 1.4 & -17.0 \\
\end{tabular}
\end{ruledtabular}
\caption{Calculated $J_1$ and $J_2$ in meV for MnO and NiO. $P_d$ is the magnetic moment of the $d$ shell
calculated from the type II antiferromagnetic phase. Values in bracket are for $P_d$ evaluated from the 
ferromagnetic ground state. ASIC$_{1/2}^*$ and ASIC$_{1}^*$ are obtained from the LDA energies evaluated at the 
ASIC density.
a) Ref. \cite{a30c},
b) Ref. \cite{podo},
c) Ref. \cite{feng},
d) Ref.  \cite{akin11},
e) Ref. \cite{temm},
f) Ref.  \cite{NiOexpt1},
g) Ref. \cite{NiOexpt2}}\label{table3}
\end{table}

Let us first focus our attention to MnO. In this case both the $J$'s are rather small and positive (antiferromagnetic 
coupling is favorite), in agreement with the Goodenough-Kanamori rules \cite{GK} and the rather low N\'eel temperature. 
Direct experimental measurements of the exchange parameters are not available and the commonly accepted
values are those obtained by fitting the magnetic susceptibility with semi-empirical methods \cite{a30c}.
Importantly this fit gives almost identical first and second nearest neighbour exchange constants. 
In contrast all the exchange functionals we have investigated offer a picture where $J_2$ is always approximately
twice as large as $J_1$. This gives us a reasonably accurate value of $J_1$ for LDA and GGA, but $J_2$ is 
overestimated by approximately a factor 2, in agreement with previous calculations \cite{akin10}.

ASIC systematically improves the LDA/GGA description, by reducing both $J_1$ and $J_2$. This is related
to the enhanced localization of the Mn $d$ electrons achieved by the ASIC, as it can be seen by comparing the
Mn $d$ magnetic moments ($P_d$) calculated for different functionals (see table \ref{table3}). 
Thus ASIC$_1$, which provides the largest magnetic moment, gives also $J$'s in closer agreement with
the experimental values, while ASIC$_{1/2}$ is not very different from LDA. 

Importantly for half-filling, as in MnO, the ASIC scheme for occupied states is fundamentally analogous to 
the LDA+$U$ method, with the advantage that the $U$ parameter does not need to be evaluated. 
Finally, at variance with KNiF$_3$, it does not seem that a portion of exact exchange is strictly necessary
in this case. Hartree-Fock \cite{podo} results in a dramatic underestimation of the $J$ parameters, while
B3LYP \cite{feng} is essentially very similar to ASIC$_1$. Curiously the best results available in the 
literature \cite{podo} are obtained with the PBE0 functional \cite{PBE0}, which combines 25\% of
exact-exchange with GGA.

The situation for NiO is rather different. The experimentally available data \cite{NiOexpt1,NiOexpt2} show 
antiferromagnetic nearest neighbour and ferromagnetic second nearest neighbour exchange parameters. 
The magnitude is also rather different with $|J_2|>10\:|J_1|$. Standard local functionals (LDA and GGA)
fail badly and overestimate both the $J$'s by more than a factor 5. ASIC in general reduces the exchange 
constants and drastically improves the agreement between theory and experiments.  In particular 
ASIC$_1$ gives exchange parameters only about twice as large as those measured experimentally.

\begin{figure}[htb]
\begin{center}
\includegraphics[width=0.45\textwidth,angle=0.0]{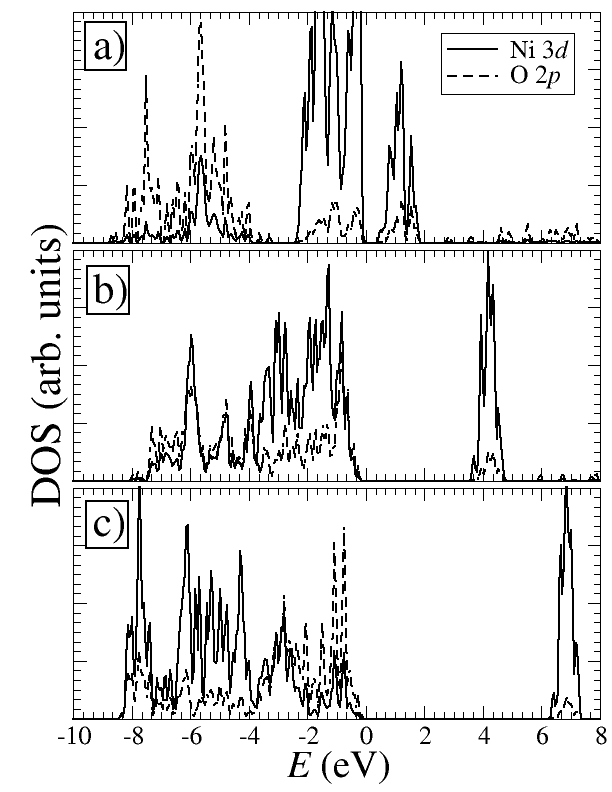}
\end{center}
\caption[] {Calculated orbital resolved DOS for type II anti-ferromagnetic NiO obtained with a) LDA, b) ASIC$_{1/2}$ 
and c) ASIC$_1$. The valence band top is aligned at 0~eV.} \label{Fig8}
\end{figure}

A better understanding can be obtained by looking at the orbital-resolved DOS for the Ni $d$ and the O $p$ orbitals 
(figure \ref{Fig8}) as calculated from LDA and ASIC. 
There are two main differences between the LDA and the ASIC results. First there is an increase of the fundamental
band-gap from 0.54~eV for LDA to 3.86~eV for ASIC$_{1/2}$ to 6.5~eV for ASIC$_1$. Secondly there is change in the
relative energy positioning of the Ni $d$ and O $p$ contributions to the valence band. In LDA the top of the
valence band is Ni $d$ in nature, with the O $p$ dominated part of the DOS lying between 4~eV
and 8~eV from the valence band top. ASIC corrects this feature and for ASIC$_{1/2}$ the O $p$ and Ni $d$ 
states are well mixed across the whole bandwidth. A further increase of the ASIC corrections ($\alpha=1$) leads
to a further downshift of the Ni $d$ band, whose contribution becomes largely suppressed close to the valence
band-top. Thus, increasing the portion of ASIC pushes NiO further into the charge transfer regime. 

Interestingly, although ASIC$_{1/2}$ gives the best bandstructure, the $J$'s obtained with ASIC$_1$ are in better
agreement with the experiments. This is somehow similar to what observed when hybrid functionals are put to the test.
Moreira et al. demonstrated \cite{akin11} that $J$'s in close agreement with experiments can be obtained by using 35\%
Hartree-Fock exchange in LDA. Moreover, in close analogy to the ASIC behaviour, as the fraction of exact exchange increases 
from LDA to purely Hartree-Fock, the exchange constants decrease while the band-gap gets larger. However, while the best 
$J$'s are obtained with 35\% exchange, a gap close to the experimental one is obtained with B3LYP, which in turns overestimates the $J$'s. 
This remarks the subtile interplay between exchange and correlations in describing the magnetic interaction of this complex 
material. Finally, it is worth remarking that a fully self-consistent SIC \cite{temm} seems to overcorrect the $J$'s, while still 
presenting  the erroneous separation between the Ni $d$ and O $p$ states. 

\section{Conclusions}

In conclusions the approximated expression for the ASIC total energy is put to the test of calculating exchange 
parameters for a variety of materials, where local and gradient-corrected XC functionals fail rather
badly. This has produced mixed results. On the one hand, the general bandstructure and in particular the 
valence band, is considerably improved and resembles closely data from photo-emission. On the other hand,
the exchange constants are close to experiments only for the case when the magnetism originates
from half-filled shells. For other fillings, as in the case of NiO or KNiF$_3$ the ASIC improvement over
LDA is less satisfactory, suggesting that a much better XC functional, incorporating a portion at least of 
exact exchange, is needed. Importantly ASIC seems to be affected by the same pathology of hybrid functional,
i.e. the amount of ASIC needed for correcting the $J$ is different from that needed for obtaining a good bandstructure.

\section{Acknowledgements}

This work is supported by Science Foundation of Ireland under the grant SFI05/RFP/PHY0062. Computational resources 
have been provided by the HEA IITAC project managed by the Trinity Center for High Performance Computing and by
ICHEC.


\begin{thebibliography}{99}
\bibitem{note} Note that here we use ``LDA'' also to refer to the local spin density approximation (LSDA), i.e. to
the spin unrestricted version of the theory.
\bibitem{note2}The ASIC method is not completely self-interaction free and therefore some spurious corrections
to the unoccuplied KS states are present.
\end{thebibliography}
\end{document}